\documentclass[12pt,onecolumn]{emulateapj}
\newcommand{\ct}{C$_{\rm 2}$}
\newcommand{\cth}{C$_{\rm 2}$H}
\newcommand{\fot}{{\case{1}{2}}}
\newcommand{\htwo}{H$_{\rm 2}$}
\newcommand{\nt}{N$_{\rm 2}$}
\newcommand{\slv}{{\sl v}}
\newcommand{\slvl}{{\sl v$_l$}}
\newcommand{\slvu}{{\sl v$_u$}}
\newcommand{\msv}{{\mathsf v}}
\newcommand{\sdss}{SDSS J223224.00$-$074434.2}
\newcommand{\teff}{$T_{\rm eff}$}
\newcommand{\trot}{$T_{\rm rot}$}

\slugcomment{Submitted to ApJ Nov. 25, 2007; Accepted Jan. 29, 2008}
\journalinfo{}
\shorttitle{PECULIAR DQ WHITE DWARFS}
\shortauthors{Hall and Maxwell}

\begin{document}
\title{C$_{\rm 2}$ in Peculiar DQ White Dwarfs}
\author{
Patrick B. Hall,\altaffilmark{1} 
Aaron J. Maxwell\altaffilmark{1} 
}
\altaffiltext{1}{Department of Physics and Astronomy, 
York University, Toronto, Ontario M3J 1P3, Canada}

\begin{abstract}
White dwarfs (WDs) with carbon absorption features in their optical spectra are
known as DQ WDs.  The subclass of peculiar DQ WDs are cool objects 
($T_{\rm eff}\lesssim 6000$~K) which show molecular absorption bands that have
centroid wavelengths $\sim$100--300\,\AA\ shortward of the bandheads of the 
\ct\ Swan bands.  These ``peculiar DQ bands" have been attributed 
to a hydrocarbon such as \cth.  We point out that \cth\ does not show strong
absorption bands with wavelengths matching those of the peculiar DQ bands
and neither does any other simple molecule or ion likely to be present
in a cool WD atmosphere.  \ct\ seems to be the only reasonable candidate for
producing the peculiar DQ bands.  Many characteristics of those bands
can be understood if they are pressure-shifted Swan bands.
While current models of WD atmospheres suggest that, in general,
peculiar DQ WDs do not have
higher photospheric pressures than normal DQ WDs do, that finding requires
confirmation by improved models of WD atmospheres and of the behavior of 
\ct\ at high pressures and temperatures.
If it is eventually shown that the peculiar DQ bands cannot be explained as
pressure-shifted Swan bands, the only explanation remaining would seem to be
that they arise from highly rotationally excited \ct\ ($J_{peak}\gtrsim 45$).
In either case, 
the absorption band profiles can in principle be used to constrain
the pressure and the rotational temperature of \ct\ in the line-forming regions
of normal and peculiar DQ WD atmospheres,
which will be useful for comparison with models.  Finally, we note that progress
in understanding magnetic DQ WDs may require models which simultaneously
consider magnetic fields, high pressures and rotational excitation of \ct.
\end{abstract}

\section{Introduction} \label{intro}

Carbon white dwarfs (DQ WDs) with $T_{\rm eff}\lesssim 11000$~K can show
absorption from the Swan bands of \ct.
Peculiar carbon white dwarfs (DQ PEC or DQp WDs) are relatively cool WDs
($T_{\rm eff}\lesssim 6000$~K) which exhibit bands reminiscent of the Swan
bands, but with more rounded profiles and with centroid wavelengths located
$\sim$100--300\,\AA\ (700$\pm$100~cm$^{-1}$) shortward of the bandheads of the
Swan bands (Table \ref{t1}; 
see also Figure 3 of {Schmidt}, {Bergeron}, \& {Fegley} 1995 and Figure 30 of {Bergeron}, {Ruiz}, \& {Leggett} 1997).
The origin of these ``peculiar DQ bands" or DQp bands has been a
matter of some debate.
{Liebert} \& {Dahn} (1983) first suggested they were pressure-shifted Swan bands,
but {Bergeron} {et~al.} (1994) cast doubt on that interpretation (see \S~\ref{nkT}).
Strong magnetic fields can and do distort the Swan bands in some carbon WDs, 
but the DQp WDs are not all strongly magnetic ({Schmidt} {et~al.} 1995, 1999), so another
explanation is required for their bands.
One well-received explanation was put forward by Schmidt et al. (1995),
who attributed the DQp bands to a hydrocarbon such as \cth\ rather than to \ct.
As a result, DQp WDs have also been referred to as \cth\ WDs.
However, until the origin of the bands in such objects is conclusively settled, 
we prefer to use the nomenclature of DQp WDs and DQp bands.

In this contribution we report an unsuccessful literature search for known
transitions from \cth\ and other simple molecules that correspond to the DQp
bands (\S~2).  We then revisit the issue of how \ct\ might produce the DQp
bands (\S~3).  We summarize our conclusions in \S~4.

\section{The Peculiar DQ Bands: Explanations Besides \ct} \label{other}

\subsection{Ruling Out \cth} \label{c2h}

Motivated by the abundance analysis of LHS~1126 by {Bergeron} {et~al.} (1994),
Schmidt et al. (1995) suggested that DQp WDs have relatively large amounts of
hydrogen ($\gtrsim$10\% of the helium abundance, compared to $\sim$0.1\%
in normal DQ WDs).
Under such conditions, neither \ct\ nor CH will be the most abundant 
carbon-hydrogen molecule.  Instead, {Schmidt} {et~al.} (1995) suggested that the 
DQp bands arise in \cth, which was generally the most abundant such 
molecule in their models.

{Schmidt} {et~al.} (1995) were not aware of theoretical or observational work on the
optical spectrum of \cth, but such work does exist.  In fact,
{Peri{\' c}}, {Peyerimhoff}, \&  {Buenker} (1992) cite 39 experimental and 13 theoretical investigations of \cth.
In particular, {Graham}, {Dismuke}, \& {Weltner} (1974) show that \cth\ 
does not produce bands similar to the \ct\ Swan bands.  The \cth\ molecule
 is electronically more like CN and has bands only at 
$\lambda<3400$\,\AA, 8670\,\AA\,$<\lambda<8750$\,AA\ and
$\lambda>9430$\,\AA. 
In no other study cited by {Peri{\' c}} {et~al.} (1992) has experimental or theoretical 
evidence been reported for bands of \cth\ in the range 4000$-$6000\,\AA.  

This fact was mentioned by {J{\o}rgensen} {et~al.} (2000), who concluded that ``the likelihood
of \cth\ being responsible for [the DQp bands] ... is very small." However, in
our opinion the wealth of previous research on \cth\ means that \cth\ has 
{\sl no} chance of being responsible for the peculiar DQ bands.
For example, even if previously unknown and unsuspected transitions of 
\cth\ matching the DQp bands exist, if \cth\ is prominent in DQp WDs we should
expect such WDs to exhibit 8710\,\AA\ band absorption from \cth\ molecules in 
the ground state.  Such absorption has not been reported in the literature, 
nor is it seen in the spectra of DQp WDs catalogued by {Harris} {et~al.} (2003),
{Dufour}, {Bergeron}, \& {Fontaine} (2005), {Koester} \& {Knist} (2006) or {Eisenstein} {et~al.} (2006).

\begin{deluxetable}{cccc}
\tablewidth{0pt}
\tablecaption{Swan and DQp Band System Parameters\label{t1}}
\tablehead{
\colhead{Swan Band} & \colhead{Swan Band} &
\colhead{DQp Band} & \colhead{$\Delta E$,} \\
\colhead{$\Delta\msv$} & \colhead{Bandhead, \AA} &
\colhead{Centroid, \AA} & \colhead{cm$^{-1}$}
}
\startdata
+4	&3828	&\nodata  &\nodata \\
+3	&4083	&\nodata  &\nodata \\
+2	&4382	&4280  &+549 \\
+1	&4737	&4575  &+748 \\
\phm{+}0 &5165	&5000  &+639 \\
$-$1	&5636	&5400  &+775 \\
$-$2	&6191	&5900  &+797 \\
$-$3	&6857	&\nodata  &\nodata \\
$-$4	&7670	&\nodata  &\nodata \\
\enddata
\end{deluxetable}

\subsection{Ruling Out Other Molecules}\label{else}

{Schmidt} {et~al.} (1995) explored molecule formation in atmospheres with an H/He abundance
ratio of 0.1 and C/He abundance ratios down to $10^{-5}$.  Depending on the
temperature and pressure, they found that a number of simple hydrocarbon
molecules could be most abundant, such as CH, \cth, \ct\htwo\ or C\htwo.
We have 
searched the literature extensively
and have ruled out these and other molecules
as being responsible for the DQp bands, either because the molecules have no
permitted optical transitions or because they have ones whose wavelengths do
not match those of the DQp bands.  The following molecules and ions
are ruled out by the following references:
C$_2^+$ ({Maier} \& {R{\"o}sslein} 1988),  
CH ({Herzberg} \& {Johns} 1969),
C\htwo\ ({Herzberg} \& {Shoosmith} 1959),
CH$_3$ ({Herzberg} \& {Johns} 1966),
CH$_4$ ({Giver} 1978),
\ct\htwo\ ({Lundberg} {et~al.} 1993),  
\cth$_3$ ({Pibel} {et~al.} 1999; {Pushkarsky} {et~al.} 2001; {Shahu} {et~al.} 2002), 
\cth$_4$ ({Adel} \& {Slipher} 1934), 
\cth$_6$ ({Dick} \& {Fink} 1977), 
C$_3$ ({Fix} 1976),
C$_3$H ({Ding} {et~al.} 2001),
C$_5$ ({Hanrath} \& {Peyerimhoff} 2001),
C$_5$H, C$_7$H and C$_9$H (Ding {et~al.} 2002).
Lastly, since it is possible that oxygen and even nitrogen as well as carbon
could be present in the atmospheres of cool WDs 
({Hansen} \& {Liebert} 2003; {J{\o}rgensen} {et~al.} 2000), we have ruled out the following
molecules containing O or N as potential explanations for the DQp bands:
HCO ({Dixon} 1969),
C$_2$O ({Choi} {et~al.} 1998),
CO ({Herzberg} 1950),
O$_2$ ({Herzberg} 1950),
OH ({Dieke} \& {Crosswhite} 1962),
H$_2$O ({Herzberg} 1966),
CN ({Herzberg} 1950) and
HCN (Schwenzer {et~al.} 1974).

A few molecules deserve special note.

We rule out absorption from metastable He$_{\rm 2}$, as suggested by
{Aslan} \& {Bues} (1999).  He$_{\rm 2}$ does produce three broad optical absorption bands,
but their wavelengths do not match those of the peculiar DQ bands (Brooks \& Hunt 1988).
Similarly, HeH (helium hydride) produces five optical absorption bands, 
but not at the right wavelengths to explain the DQp bands (Ketterle, Figger, \& Walther 1985).

C$_2^-$ ({Herzberg} \& {Lagerqvist} 1968; {Lineberger} \& {Patterson} 1972)
has a bandhead at 5416 \AA, close to an observed DQp band centroid at 5400 \AA.
However, that C$_2^-$ band is blue-shaded, not symmetric like the DQp band.
Furthermore, the neighboring C$_2^-$ bandheads are at 4902 \AA\ and 5985 \AA,
respectively, which are not good fits to the neighboring DQp bands.

The rhombic C$_4$ isomer is predicted to produce an electronic band of
comparable strength to the Swan bands ({Muhlhauser} {et~al.} 2000; {Kokkin}, {Bacskay}, \& {Schmidt} 2007) in the vicinity of
5450-5800\,\AA\ via absorption from the $X^1A_g$ ground state to the $^1B_{1u}$
excited state (Masso {et~al.} 2006).  However, the band has not been observed
experimentally and related vibronic band predictions have not been made.
The linear isomer of C$_4$ produces UV/optical bands at 3340-3790\,\AA,
but they are about 50 times weaker than the Swan bands ({Linnartz} {et~al.} 2000; {Muhlhauser} {et~al.} 2000).  
No information is available about optical bands of the branched C$_4$ isomer,
which is thought to be a nearly planar tetrahedron ({Hochlaf}, {Nicolas}, \& {Poisson} 2007).
C$_4$ would be a long-shot explanation for the DQp bands, but the existing data
do not rule it out.  Absorption from all three C$_4$ isomers would be expected
in that case.  Bands from linear C$_4$ are not seen in one DQp WD with the
required spectral coverage ({Hintzen} 1986), but the bands are expected to be weak.

Finally, for completeness we note that isotopically shifted Swan bands cannot
explain the DQp bands: $^{13}$C$^{12}$C and $^{13}$C$^{13}$C have bands shifted
about 8\,\AA\ and 16\,\AA\ {\em longward} in wavelength, respectively, 
from those of \ct\ ({King} \& {Birge} 1929; {Swings} 1943).

\subsection{An Unknown Diatomic Molecule?} \label{unknown}

If we treat the DQp bands as being from an unknown diatomic molecule,
we can infer some approximate characteristics of the two electronic
states involved.  We assume that the DQp centroid wavelengths given in
{Schmidt} {et~al.} (1999) are the wavelengths of the (\slvl,\slvu) vibrational bands
(0,2), (0,1), (0,0), (1,0) and (2,0) of the same electronic transition,
where \slvl\ and \slvu\ are the initial and final vibrational quantum numbers.
The wavenumber of the (\slvl,\slvu) vibronic transition can be approximated as
\begin{equation} \label{os}
\omega_{lu}= \Delta E/hc
+\omega_u(\msv_u+\fot)
-\omega_ux_u(\msv_u+\fot)^2
-\omega_l(\msv_l+\fot)
+\omega_lx_l(\msv_l+\fot)^2
\end{equation}
where $\Delta E=E_u-E_l$ is the (unknown) energy difference between the 
electronic states involved, $h$ is Planck's constant, $c$ is lightspeed 
and $\omega_u$, $\omega_ux_u$, $\omega_l$ and $\omega_lx_l$
are constants to be determined ({Herzberg} 1950).
A pure harmonic oscillator potential would have $\omega x$=0, while for
realistic potentials there are likely to be additional higher-order
terms; nonetheless, the approximation in Equation \ref{os} is useful.

Five peculiar DQ bands have been observed, so we can use those five
$\omega_{lu}$ values with their (assumed) associated \slvl\ and \slvu\ to
solve for the five unknowns.
Table \ref{t2}, where they are compared with the corresponding 
parameters for the Swan bands and for the optical bands of C$_2^-$.
Taking these coefficients at face value, the energy difference between the DQp
electronic states is 1.8\% larger than between the Swan band electronic states.
The DQp $\omega x$ values are large relative to the $\omega$ values,
indicating much more anharmonic potentials than for either state involved
with the Swan bands.  
The smaller $\omega_l$ value for the DQp bands indicates a lower state with
a potential energy curve broader than that of the Swan band lower state,
and the larger $\omega_u$ value for the DQp bands indicates a upper state with
a potential energy curve narrower than that of the Swan band upper state.
Last but not least, the negative value of $\omega_lx_l$ is extremely unusual
({Herzberg} 1950).  We can find only one diatomic molecule containing H, C or O with
any transition characterized by a negative $\omega x$; namely, CsH ({Huber} \& {Herzberg} 1979).

Thus, the characteristics of the DQp bands argue against attribution of
the bands to an unidentified diatomic molecule. 
In deriving those characteristics, however, we did equate the band centroids
with the wavelengths of transitions involving a $\msv=0$ vibrational level.
If that assumption is not correct, then the characteristics of the
energy states involved may be less unusual.  
Since the long-wavelength edges of the DQp bands have wavelengths consistent
with the $\msv=0$ bandheads of the Swan system,
it seems plausible that the DQp bands are the Swan bands,
a possibility we now consider.

\begin{deluxetable}{lrrrrr}
\tablewidth{0pt}
\tablecaption{Band System Constants\label{t2}}
\tablehead{
\colhead{Band System} & \colhead{$\Delta E$} &
\colhead{$\omega_l$} & \colhead{$\omega_lx_l$} &
\colhead{$\omega_u$} & \colhead{$\omega_ux_u$} 
}
\startdata
DQp & 19656\phm{.00} & 1395.3\phm{0} & $-$44.36 & 2201.9\phm{0} & 173.4\phm{0} \\ 
Swan    & 19306.26 & 1641.33 & 11.65 & 1788.22 & 16.46 \\
C$_2^-$ $(X^2\Sigma_g^+,B^2\Sigma_u^+)$                                                 & 18390.88 & 1781.04 & 11.58 & 1968.73 & 14.43 \\
\enddata
\end{deluxetable}

\section{The Peculiar DQ Bands: \ct\ as an Explanation} \label{c2}

If the references cited in the sections 2.1 and 2.2
are correct (and for them to be wrong would require laboratory spectroscopy
to have missed  transitions as strong as the \ct\ Swan bands), then 
the similarity of the bands in peculiar DQ WDs to those in normal DQ WDs
suggests that \ct\ produces the peculiar DQ bands.
In this section we use observations and basic molecular physics
to consider how \ct\ might be producing the peculiar DQ bands.
We begin by summarizing the salient features of the \ct\ molecule
and the Swan bands.

\subsection{Basic Physics of \ct\ and the Swan Bands} \label{basics}

The electron configuration of the \ct\ ground state ($X^1\Sigma^+_g$)
can be approximated as $\sigma^{*2}_{u2s} \pi^4_{u2p}$
(omitting the inner orbitals $\sigma^2_{g1s} \sigma^{*2}_{u1s} \sigma^2_{g2s}$),
while that of the Swan band lower ($a^3\Pi_u$) and upper ($d^3\Pi_g$) states are
$\sigma^{*2}_{u2s}\pi^3_{u2p}\sigma_{g2p}$
and
$\sigma^*_{u2s}\pi^3_{u2p}\sigma^2_{g2p}$
respectively (Mulliken 1939).
That is, the Swan bands are produced by vibronic transitions where an electron
moves from a vibrational level in the $\sigma^*_{2s}$ electronic orbital
to one in the $\sigma_{2p}$ orbital.  
(Unstarred orbitals are bonding orbitals,
while starred orbitals are antibonding.  The subscripts $g$ and $u$ indicate
symmetry and antisymmetry, respectively, of the electron wavefunctions with
respect to reflection through the center of the molecule.)
Because of the transfer of an electron from an antibonding to a bonding orbital,
Swan band absorption puts the \ct\ molecule in an upper state where it has a
3.7\% shorter equilibrium internuclear separation $r_e$
than it did in the lower state (Mulliken 1939; {Prasad} \& {Bernath} 1994).

The Swan bands are made up of many individual rovibronic 
transitions (transitions with simultaneous changes in rotational quantum 
number $J$, vibrational quantum number \slv, and electronic state).
Each individual Swan band is
characterized by a particular $\Delta\msv$=\slvl$-$\slvu\ and
consists of overlapping rovibronic bands with different (\slvl,\slvu).
Each rovibronic band consists of all possible rotational transitions for
a particular vibronic transition (characterized by a given $\Delta\msv$ and
\slvl).  For example, the
$\Delta\msv$=0 Swan band contains rovibronic bands (0,0), (1,1), etc.

Each rovibronic band consists of a $P$ branch ($J\rightarrow J-1$)
and an $R$ branch ($J\rightarrow J+1$).
For the Swan bands with $|\Delta\msv|\leq 2$, 
as $J=J_l$ increases with each rovibronic band the wavelengths
of the corresponding $R$ branch transitions continually decrease.
In contrast, with increasing $J$ the wavelengths of the corresponding $P$
branch transitions first increase, then reach a maximum (forming the bandhead
of the rovibronic band), and then continually decrease.

The Swan bands are usually said to be shaded to the blue, meaning
that rovibronic bands with larger \slvl\ have bandheads at shorter wavelengths.
However, this is strictly true only for $\msv_l\leq 7+2\Delta\msv$:
at sufficiently large \slvl\ the 
rovibronic bandheads shift to longer wavelengths with increasing \slvl.

The relative absorption strength of a particular Swan band rovibronic
transition ($J_l\rightarrow J_u$, $\msv_l\rightarrow\msv_u$) depends on 
the population of the ($J_l,\msv_l$) level and the square of the overlap
integral $\int \psi^*_l\psi_u^{}\,d\vec{r}$, known as the Franck-Condon factor.
Simply put, the more the maxima of the electron probability densities
in the lower and upper rovibronic levels match up,
the greater the probability of a transition between them.

\subsection{Pressure-Shifted Swan Bands?} \label{nkT}

The high gravity of white dwarfs causes gravitational separation of elements.
If there is substantial helium and hydrogen in a WD, the WD will have
a hydrogen envelope atop a helium layer atop a carbon layer, etc.
As white dwarfs cool below $T_{\rm eff}\lesssim 16000$~K, the 
hydrogen and helium layers 
(or the helium layer alone, if hydrogen is absent)
can become deeply convective if they are not too thick.
Thus, some cool WDs have helium-dominated atmospheres with trace amounts of
carbon ($10^{-7}-10^{-2}$; {Koester} {et~al.} 1982;
{Pelletier} {et~al.} 1986) as well as
hydrogen ($10^{-6}-10^{-1}$; {Schmidt} {et~al.} 1995).
Because He is predominantly neutral at such temperatures, the atmosphere of
a cool He-rich WD is highly transparent in the optical and the photosphere is
located relatively deep within the star, at high pressure
($P\simeq 10^{11}$ dyne cm$^{-2}$ $=10$~GPa;
{J{\o}rgensen} {et~al.} 2000).
The dominant source of opacity in such atmospheres is He$^-$ 
({J{\o}rgensen} {et~al.} 2000),
with trace heavy elements such as carbon providing the donor electrons.
A lower metal abundance should therefore decrease the atmospheric opacity,
shifting the photosphere downwards, to higher pressure.
As $T_{\rm eff}$ decreases in DQ WDs, the carbon abundance is observed to
decrease due to a decreasing efficiency of convective carbon dredge-up
({Dufour} {et~al.} 2005; {Koester} \& {Knist} 2006).  It is therefore at least plausible that the DQp bands
could be Swan bands affected by high pressure.

The gas pressure in a WD is governed by the hydrostatic equation
$dP_g/d\tau = g\rho/k$, where $\tau$ is the optical depth, $g$ the acceleration
due to gravity, $\rho$ the density and $k$ the total linear absorption
coefficient from all sources of opacity ({Allard} \& {Wehrse} 1990).  
If DQp bands are generated by \ct\ at high pressure, 
then at $\tau\sim 1$ in a DQp WD atmosphere $k$ must be larger,
or $g$ or $\rho$ smaller, than at $\tau\sim 1$ in a DQ WD atmosphere.
For a cool, He-rich WD to become a DQp WD, a low metal abundance might be
needed, since that would reduce the availability of donor electrons and reduce
the He$^-$ opacity, thus shifting the photosphere to higher pressures.  
Alternatively, a larger value of $g$ might be required, due to DQp WDs having
typically higher masses than DQ WDs.  However, there is no evidence that the
latter is the case: {Bergeron}, {Leggett}, \& {Ruiz} (2001) do not find systematically higher masses for DQp WDs
as compared to DQ WDs, and there is some evidence that a subset of {\em normal}
DQ WDs have systematicallly large masses ({Dufour} {et~al.} 2005).  

Pressure shifting of the Swan bands as an explanation for the DQp bands was 
originally suggested by {Liebert} \& {Dahn} (1983) but seemingly
ruled out by {Bergeron} {et~al.} (1994).  In the latter paper, the photospheric pressure in
a model for the DQp WD LHS 1126 was found to be three times smaller than the 
photospheric pressures in the coolest normal DQ WDs analyzed by {Wegner} \& {Yackovich} (1984).  
However, numerous inadequacies in current models of cool white dwarf 
atmospheres have recently come to light (e.g., {Kowalski}, {Saumon}, \& {Mazevet} 2005; {Dufour} {et~al.} 2007a).
For LHS 1126 specifically,
the large H/He ratio of 10$^{-0.8}$ found by {Bergeron} {et~al.} (1994) was 
shown by {Wolff}, {Koester}, \& {Liebert} (2002) to grossly overpredict the Ly$\alpha$ absorption 
strength in that object, which is most consistent with H/He=10$^{-4.5}$.
Furthermore, neither model correctly predicts the mid-infrared fluxes of LHS
1126 recently reported by {Kilic} {et~al.} (2006).
No current model can satisfactorily explain the entire spectral energy 
distribution of the DQp WD LHS 1126.  Therefore, the conclusion that it has 
a lower photospheric pressure than some normal DQ WDs, and thus that the DQp 
bands cannot be pressure-shifted Swan bands, may have been premature.

As improved models of cool, He-rich WD atmospheres become available,
it will be worth refitting well-observed DQ and DQp WDs
to determine whether the DQp bands could be pressure-shifted Swan bands.
For example, {Homeier}, {Allard}, \& {Allard} (2007) have shown that at sufficiently high pressure,
collisions of Na atoms with \htwo\ or He
can shift \ion{Na}{1} absorption shortwards as well as broaden it.  It is worth 
investigating if collisions of \ct\ with \htwo\ or He could produce a similar 
effect and, if so, at what pressures.
(A 50~\AA\ shortward shift of the Swan bands at a pressure of 
10$^{10}$ dynes cm$^{-2}$ was reported by {Bues} (1999),
but no details of the calculation were given.)

Theoretically,
Lin (1973) has shown that pressure can shift vibronic bands either shortward
or longward in wavelength, that the energy shift is identical for each band, and
that such bands are likely to appear increasingly Gaussian at higher pressure.
The DQp bands are in fact shifted by from the Swan bands by roughly identical
energies (+700$\pm$100~cm$^{-1}$) and appear roughly Gaussian,
so the DQp bands do possess some features
of pressure-shifted bands.

Experimentally,
\htwo\ and \nt\ immersed in dense, high-pressure
He show an increase in vibrational energy level spacing (Loubeyre, LeToullec, \& Pinceaux 1992; Scheerboom, Michels, \& Schouten 1996).
The increase is due simply to compression of the molecule: shortening of the
intramolecular bond leads to more energy being required to excite vibrations.
Unfortunately, even assuming the same is true for \ct, no rovibronic bands 
which could be used to measure such an increase appear within the DQp bands.
Neither is experimental information available on vibronic band shifts for 
\htwo, \nt\ or \ct\ in He.  However, in mixtures of 1\% \nt\ in low-temperature
Ar, Kr or Xe, at least one \nt\ vibronic band is seen to shift to {\em longer}
wavelengths as a function of increasing pressure (Semling {et~al.} 1997).
The upper state involved in those \nt\ transitions has a larger $r_e$ than
the lower state, so it is likely that the upper state undergoes greater
compression (Lin 1973), making it more tightly bound and reducing the energy
difference between it and the lower state.
If that is the case, the Swan bands may behave oppositely
(vibronic bands shifted shortwards), since the lower state involved
is larger and thus likely to undergo greater compression under pressure.

Empirically, the DQp bands exhibit some characteristics consistent
with being pressure-shifted Swan bands where the lower state involved
is more affected by pressure than the upper state.
As mentioned in the previous paragraph,
the DQp bands show no evidence for bandheads.
For the bandheads formed by the rotational $P$ branches to
disappear without bandheads appearing in the $R$ branches, the rotational
constants in the (shifted) lower and upper Swan band states must be nearly
equal: $B_l\simeq B_u$ ({Herzberg} 1950).  Since 
$B\propto I^{-1}$, where $I\propto r_e^2$ 
is the moment of inertia perpendicular to the internuclear axis,
nearly equal values of $B$ require nearly equal internuclear separations
in the two states.  
That requirement is consistent with the Swan band
lower state being more affected by pressure than the upper state.
Another expected consequence of more nearly equal internuclear separations is a
shift in the peak absorption within each Swan band to shorter wavelengths due to
smaller (larger) Franck-Condon factors for levels with small (large) 
\slvl.  This narrowing of the Condon parabola ({Herzberg} 1950; {Cooper} 1975) may also reduce
the overall intensities of the $|\Delta\msv|\geq2$ bands in DQp WDs relative
to DQ WDs (see \S~\ref{MAG}).

In summary, it seems plausible that high pressures and temperatures could 
distort the Swan bands in such a manner as to produce the DQp bands.
If detailed calculations or laboratory experiments show that 
pressure-shifted Swan bands do indeed resemble the DQp bands, the 
claim that DQp WDs do not have higher photospheric pressures
than DQ WDs will need to be checked with better models of cool WD atmospheres.

\subsection{Absorption from Vibrationally or Rotationally Excited \ct?}\label{rovib}

Generally speaking, absorption from rotationally or vibrationally excited
levels of \ct\ appears at shorter wavelengths than the Swan bandheads.  Thus,
as pointed out by {Schmidt} {et~al.} (1995), such absorption might explain the DQp bands.
We consider these possibilities in detail here.

Vibrational excitation of \ct\ by itself will not suffice.  
For Swan bands with $\Delta\msv\leq 0$,
absorption from $\msv_l\approx 7$ can match the DQp band centroids.
However, for Swan bands with $\Delta\msv>0$, the rovibronic bandheads 
never reach wavelengths as short as the DQp band centroids ({Phillips} \& {Davis} 1968).
Moreover, a very high temperature of $T_{vib}\gtrsim17000$~K would be required
to populate the \slvl=7 level, as it is located $\sim$12000 cm$^{-1}$ above the
\slvl=0 level.
Furthermore, 
the $\Delta\msv=0$ Swan band has very small Franck-Condon factors for \slvl$>$3,
so vibrationally excited \ct\ should produce a very weak $\Delta\msv=0$ band.
Lastly, in thermodynamic equilibrium, even if levels with $\msv_l\approx 7$
were populated, levels with \slvl$<$0 should be equally populated.
That would yield absorption profiles {\em extended} to shorter wavelengths,
but not {\em shifted} to shorter wavelengths.

Rotational excitation of \ct\ could explain the DQp bands, since 
the $2J+1$ degeneracy of rotational levels yields level populations
which peak at a nonzero $J$ value.
Since the DQp band profiles are quite smooth even at the location of Swan band
rovibronic bandheads, the wavelength shift for the $J_{peak}$ level
in both the $P$ and $R$ branches
must be larger than the wavelength separation of rovibronic bandheads.  Such a
shift requires $J_{peak}\gtrsim 45$ ({Phillips} \& {Davis} 1968), corresponding to energies
$\gtrsim 3600$~cm$^{-1}$ above the $J=0$ level, or $T_{rot}\gtrsim 5200$~K.
Absorption from rotationally excited \ct\ could be a plausible explanation
for the DQp bands if \ct\ absorption in a DQ WD occurs high in the atmosphere,
so that $T_{rot}$ is low, while the \ct\ absorption in a DQp WD occurs deep
in the atmosphere, so that $T_{rot}$ is high.
Such a scenario is broadly in agreement with the models of {Bues} (1999).

\begin{figure}
\plotone{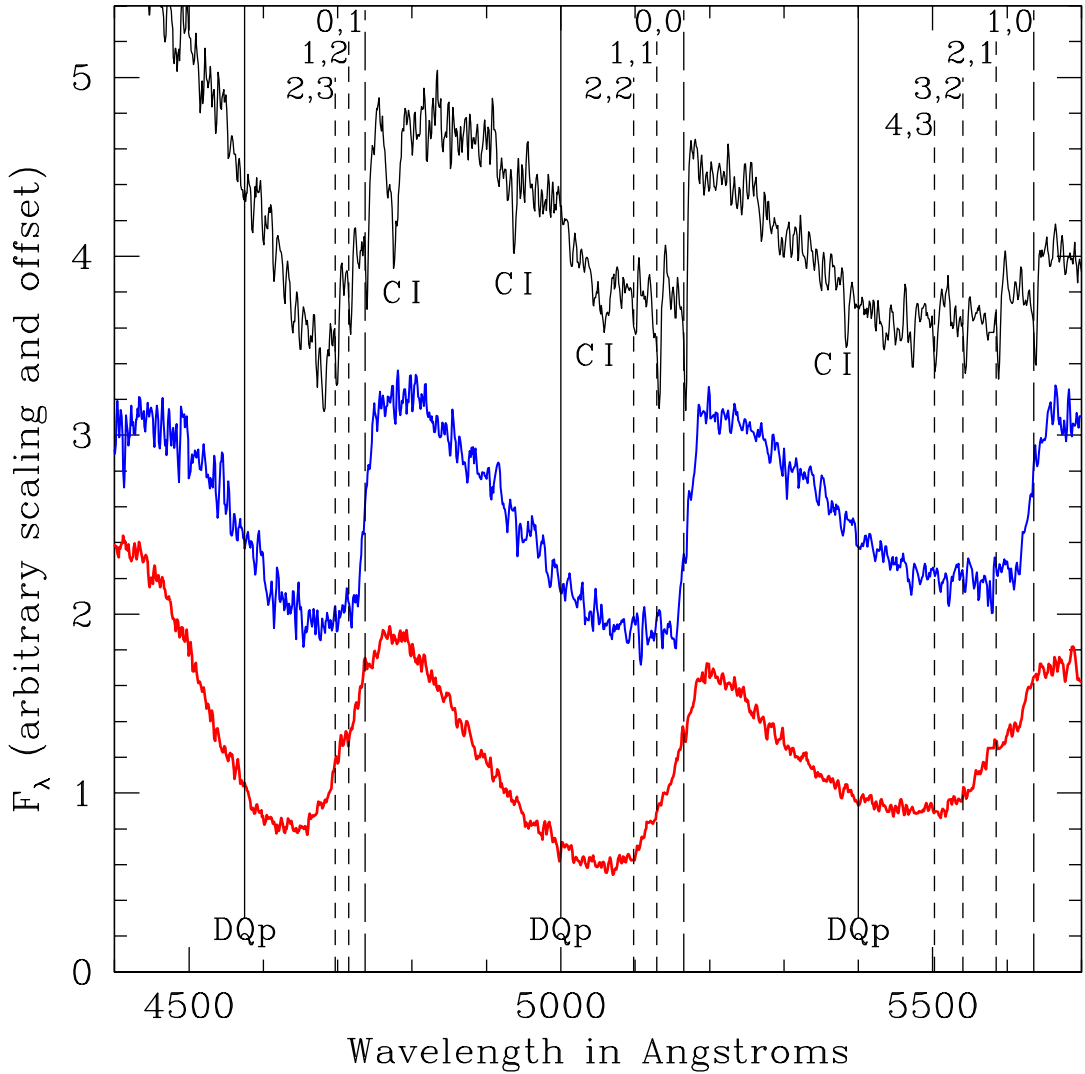}
\caption{Partial spectra of DQ and DQp WDs from the SDSS.
Solid vertical lines show the DQp band centroid wavelengths from Schmidt et al.
(1995).  Dashed lines show bandhead wavelengths of the Swan band system, while
dotted lines show rovibronic bandhead wavelengths within different Swan bands;
all lines are labeled with (\slvl,\slvu).
At the top is the DQ WD SDSS J104559.13+590448.3 (thin line),
which shows narrow absorption in the Swan bands as well as from \ion{C}{1}.
In the middle is the DQ WD SDSS J102801.74+351257.9 (thick blue line),
which has excited levels of C$_2$ sufficiently populated
that different rovibronic bands are blended together,
as well as some line broadening (pressure broadening?) which extends
the absorption profiles slightly longward of the Swan bandheads.
At the bottom is the DQp WD SDSS J223224.00$-$074434.2 (thicker red line),
whose absorption profiles show a similar amount of line broadening
along with a band centroid shift about half that seen in the DQp stars of
Schmidt et al. (1995).  Its absorption profiles are smooth but not featureless;
the slope of a profile tends to change near the wavelength of a Swan band
rovibronic bandhead (dotted lines), suggesting that the DQp bands are
modified Swan bands.
\label{fig1}}
\end{figure}

These features of Swan and DQp bands are illustrated in Figure~1,
which shows partial spectra from the Sloan Digital Sky Survey
(SDSS; {York} {et~al.} 2000; {Adelman-McCarthy} {et~al.} 2008) of two DQ WDs 
and the DQp WD \sdss\ ({Harris} {et~al.} 2003).
The existence of the latter object proves that if the DQp bands 
are shifted Swan bands, the underlying physical mechanism 
does not shift the bands by only a single, fixed energy,
as was believed to be the case at the time of the {Schmidt} {et~al.} (1995) study.
Furthermore,
there is some indication in the spectrum of \sdss\ that the rovibronic
bandheads are not completely washed out.  This suggests either a lower
pressure or a lower $T_{rot}$ in the line-forming region of \sdss\ as
compared to other DQp WDs where no sign of those bandheads remains.
Note that $J_{peak}\gtrsim 20$ is required to shift
Swan band absorption significantly away from the rovibronic bandheads,
since those bandheads are formed by the $P$ branch with $J\simeq 15$.
(For $J_{peak}\lesssim 20$, the absorption will be extended to shorter
wavelengths but the bandheads will not be shifted to shorter wavelengths.)

Detailed modeling of the bands in DQ and DQp WDs should be able to
determine whether combined rotational and vibrational excitation 
at a single $T=T_{rot}=T_{vib}$ can explain the band shapes
quantitatively as well as qualitatively.  If so, it will be possible 
to determine the excitation temperature ranges seen in DQ and DQp WDs
for comparison with models.  If not, pressure-shifted Swan bands would
seem to be the only viable explanation left for the DQp bands.

\subsection{Magnetic Considerations}\label{MAG}

It is now feasible to predict the spectra of \ct\ in the presence of magnetic
fields of $\lesssim 10^7$ Gauss ({Berdyugina}, {Berdyugin}, \&  {Piirola} 2007), though no detailed calculations
have been published for higher field strengths where the quadratic Zeeman
effect becomes important.  Such modeling is important because magnetic fields
can shift the Swan bands to shorter wavelengths ({Liebert} {et~al.} 1978),
producing bands that resemble
the (nonmagnetic) DQp bands.  In a magnetic field, the $P$ and $R$ branches of
the Swan bands become circularly polarized in opposite senses ({Berdyugina} {et~al.} 2007).
Because the $P$ and $R$ branches are not separated in wavelength, the net Swan
band circular polarization signal is weak for fields $\lesssim 10^7$ Gauss
({Schmidt} {et~al.} 2003).  However, it can reach $\sim$10\% at $\gtrsim 10^8$ Gauss
({Schmidt} {et~al.} 1999), which may be a result of quadratic Zeeman shifts.

{Schmidt} {et~al.} (1999) suggested that DQp bands show weaker quadratic Zeeman shifts than
the Swan bands do. 
However, this suggestion is not a certainty, given the possibility of a range
of magnetic field strengths (as well as geometries and orientations) combined
with a range of DQp band centroid shifts due to pressure, rotational excitation,
or both.  For example, it may be that a peculiar DQ WD with an 
intrinsically moderate band centroid shift {\em and} a moderate magnetic field
can have a spectrum similar to a normal DQ WD with a strong magnetic field.
Specifically, {Schmidt} {et~al.} (1999) suggest that LP~790$-$29 is a magnetic DQ and LHS~2229
a magnetic DQp.  We suggest that {\em both} might be DQp WDs, with
LHS~2229 having the stronger magnetic field of the pair
but LP~790$-$29 having a much larger intrinsic DQp band centroid shift.
The relative field strength estimate is based on 
the maximum change in circular polarization between wavelengths
outside and inside the Swan bands
being larger in LHS~2229 than in LP~790$-$29 (25\% vs. 10\%);
of course, the change in circular polarization could be affected by
magnetic field geometry and orientation as well.
The intrinsic DQp band centroid shift estimate assumes that
the pronounced weakness of the $|\Delta\msv|\geq 2$ bands in
LP~790$-$29 is due to smaller Franck-Condon factors between energy states 
with reduced internuclear separations (\S~\ref{nkT}).  Similar conclusions 
might apply to the similar WDs from {Schmidt} {et~al.} (2003): the band centroid shifts in
SDSS J133359.86+001654.8 are larger than in LHS~2229,
and those in SDSS J111341.33+014641.7 are larger than in
LP~790$-$29, but each SDSS object is {\em less} polarized than its 
non-SDSS counterpart.
We suggest that those two SDSS objects have considerably larger intrinsic DQp
band centroid shifts but somewhat smaller magnetic shifts than their respective
non-SDSS counterparts, leading to greater overall absorption band shifts.

Progress on understanding magnetic DQ WDs may therefore require calculations
which simultaneously incorporate the effects of pressure, rotational (and
vibrational) excitation and magnetic field strengths above 10$^7$ Gauss.

\section{Conclusions} \label{end}

We have shown that the molecular absorption bands in peculiar DQ WDs are very
unlikely to be produced in any molecule other than \ct\ (\S~2).  We have
considered how the DQp bands can be explained as modified Swan bands
(\S~3) and conclude that they could either be 
pressure-shifted Swan bands (as originally suggested by {Liebert} \& {Dahn} 1983)
or Swan band absorption from highly rotationally excited levels, or both.
Regardless, a continuum of increasingly shifted and reshaped bands is
expected to be observed in DQp WDs, in agreement with recent observations
(\S~\ref{rovib}).

One implication of our result is that the putative absence of WDs with
$T_{\rm eff}\lesssim 6000$~K which exhibit molecular carbon absorption 
({Dufour} {et~al.} 2005) is not real --- the DQp WDs are such objects.
Another implication is that understanding the spectra of magnetic DQ WDs
may require models which simultaneously account for the effects of pressure,
rotational excitation and magnetic fields on \ct\ absorption.

Theoretical and experimental investigations of the properties of \ct\ in
high-pressure, high-temperature helium are needed to determine if 
pressure-shifted Swan bands resemble the DQp bands.  If they do not, then
the only remaining explanation for the DQp bands seems to be that they arise
in rotationally excited \ct\ ($J_{peak}\gtrsim 45$). 
That scenario would require a higher \trot\ in DQp WDs than in DQ WDs,
perhaps due to \ct\ absorption arising deeper in the photosphere in DQp WDs
({Bues} 1999).

Even if pressure-shifted Swan bands are found to resemble the DQp bands,
existing models of WD atmospheres suggest that DQp WDs do not have higher 
photospheric pressures than DQ WDs and thus that the peculiar DQ bands are not 
pressure-shifted Swan bands.
Nonetheless, since there are known shortcomings in existing models of cool,
He-rich WD stellar atmospheres, that tentative conclusion needs to be
revisited when improved models become available.\footnote{More speculatively,
it would also be worth investigating the properties of carbon-atmosphere WDs
({Dufour} {et~al.} 2007b) at low \teff, to see if they have high photospheric pressures.
Although {Dufour} {et~al.} (2007b) state that no carbon-atmosphere WD is known at
\teff$<15,000$~K, carbon-atmosphere WDs were not thought to exist at all
before their recent discovery.
We cannot be confident that all carbon-atmosphere WDs
develop a helium atmosphere as they cool until the origin of such objects
is better understood.  If predictions can be made of the spectra of
carbon-atmosphere WDs at low \teff, the SDSS provides a large database in
which to search for objects with such spectra.}
If DQp WDs do have higher photospheric pressures than DQ WDs, 
the reason may be that low metal abundances reduce the He$^-$ opacity by
restricting the availability of donor electrons.

Whichever explanation of the DQp bands is eventually confirmed, with accurate
theoretical models it should be possible to use \ct\ absorption band profiles
as a sort of barometer
to constrain both the pressure and the rotational temperature of \ct\ in the
line-forming regions of DQ and DQp WD atmospheres.  For example,
examination of Swan band fits to DQ WD absorption profiles shows that the
observed profiles are often more rounded than predicted ({Dufour} {et~al.} 2005; {Koester} \& {Knist} 2006).
Thus, the observed profiles may provide evidence for an increasing $J_{peak}$ 
or an increasing pressure shift with decreasing $T_{\rm eff}$ even among
normal DQ WDs.  However, these discrepancies could simply be due 
to imprecise knowledge of Swan band absorption parameters
or to inexact treatment of the WD atmospheres or of \ct\ itself.  For example,
{Dufour} {et~al.} (2005) calculate \ct\ line profiles using the impact approximation,
which is known to break down in dense DQ WD atmospheres where the mean
distance between particles is comparable to the particle size ({Koester}, {Weidemann}, \& {Zeidler} 1982).

Observationally, more and better parallax values for DQ and DQp WDs would
provide information on the range of masses seen in each type of WD and help
determine what contribution different values of $g$ make to the DQp phenomenon.
It might also be useful to 
search for and study other \ct\ bands in normal and peculiar DQ WDs.
Whatever physical effect transforms the Swan bands into the
DQp bands, its effect on other \ct\ energy states, and thus other \ct\ bands,
may be different and may serve as a clue to its origin.

\acknowledgments
We dedicate this paper to Dr. Ralph Nicholls, who passed away in January 2008
after working at York University on \ct\ and many other molecules
since before either of this paper's authors were born.
We thank H. Harris and G. Schmidt for discussions and
M. Horbatsch for refusing to believe that the spectrum of \cth\ was unknown.
PBH acknowledges support from York University and NSERC.
This research would not have been feasible without the collections of the
York University Libraries and the Ontario Council of University Libraries.

\end{document}